\documentclass[twocolumn,english,aps,showpacs,floatfix,here]{revtex4}
\usepackage[T1]{fontenc}
\usepackage[latin1]{inputenc}
\usepackage{color}
\usepackage{amssymb}

\makeatletter

\providecommand{\tabularnewline}{\\}

\usepackage{graphicx}
\usepackage{amssymb}

\usepackage{graphicx}
\usepackage{amssymb}

\usepackage{float}
\usepackage{graphicx}
\usepackage{amssymb}
\usepackage{color} 

\usepackage{babel}

\usepackage{babel}

\usepackage{babel}
\makeatother

\begin{document}

\title{Effective-interaction approach to the many-boson problem}

\author{J.~Christensson$^{1}$, C.~Forssén$^{2}$, S.~\AA berg$^{1}$,
and S.M.~Reimann$^{1}$}

\affiliation{$^{1}$ Mathematical Physics, LTH, Lund University, Box 118, 22100
Lund, Sweden\\
 $^{2}$ Department of Fundamental Physics, Chalmers University of
Technology, 412 96 Göteborg, Sweden}

\begin{abstract}
We show that the convergence behavior of the many-body numerical diagonalization
scheme for strongly interacting bosons in a trap can be significantly
improved by the Lee-Suzuki method adapted from nuclear physics: One
can construct an effective interaction that acts in a space much smaller
than the original Hilbert space. In particular for short-ranged forces
and strong correlations, the method offers a good estimate of the
energy \textit{and} the excitation spectrum, at a computational cost
several orders of magnitude smaller than that required by the standard
method. 
\end{abstract}



\maketitle
The many-body problem of interacting bosons or fermions poses a continuous
challenge for quantum physics. The complexity of the quantum states
increases quickly with higher particle numbers or stronger correlations,
making it essentially impossible to solve the many-body problem exactly.
Thus, mean-field methods are often applied, being numerically much
less demanding than any attempt to diagonalize the many-body Hamiltonian.
Examples are the Gross-Pitaevskii approach for trapped bosons~\cite{gross-pitaevskii1961,leggett2001,pitaevskii-stringari2003},
or the celebrated Kohn-Sham equations~\cite{kohn-sham1965,kohn-nobel}
for fermions, as often used in the local (spin) density approximation.
However, these methods can treat correlations only in an approximate
way, and are proven insufficient for strong interactions between the
particles. Quantum Monte Carlo calculations provide an alternative
approach in many cases (see for example, the review by Harju~\cite{harju2005}),
but have other drawbacks, such as limited accessibility of the excitation
spectrum, a problem shared by the other methods. The development of
alternative diagonalization methods such as, e.g., coupled cluster
methods \cite{cederbaum-coupled-cluster} is an urgent issue.

Facing the above problems with strong interactions, one quickly realizes
that often, the only method of choice is the straightforward numerical
diagonalization of the many-body Hamiltonian. In fact, this has been
tried for a large variety of physical problems, ranging from nuclear
structure (see for example, the review by Caurier \textit{et al.}~\cite{caurier2005})
and quantum chemistry to artificially made quantum systems such as
metallic clusters~\cite{koskinen1994} and quantum dots~\cite{reimann2002}.
However, with increasing particle number or stronger interactions
between the particles, the number of basis states needed for an accurate
description of the many-body quantum system increases beyond computational
reach. Truncations of Hilbert space become necessary -- but often,
for the reduced basis, the results are too inaccurate~\cite{alon2004}.

In nuclear physics, a significant step forward has been achieved by
applying the Lee-Suzuki method~\cite{Lee-Suzuki,unitary-operator},
that prescribes unitary transformations on operators (e.g. the Hamiltonian)
to obtain effective operators within the reduced basis space. This
method has been successfully used in so called no-core shell model
calculations, for example, where nuclear systems with $\sim$12 fermions
have been studied~\cite{ncsm}.

Since the experimental realization of Bose-Einstein condensation with
cold, trapped atoms, much interest turned to the physics of harmonically
confined many-boson systems. Although the trapping potentials today
confine thousands of bosons, the few-body regime does not seem to
be impossible to reach, having in mind also the recent advances with
optical lattices \cite{optical-lattices}. Increasing technological
expertise with Bose-Einstein condensates {}``on chips'' \cite{atoms-on-chips},
together with the newly emerging research area of {}``atomtronics''~\cite{atomtronics},
makes the need for further theoretical developments for a description
of cold-atom gases with strong correlations an urgent issue.

In this Letter, we apply the Lee-Suzuki method -- to the best of our
knowledge, for the first time -- to a system of harmonically trapped
spinless \textit{bosons} with short-ranged repulsive interactions.
The fact that the Lee-Suzuki method works very well for the strong
short-ranged interactions between the nucleons encourages its application
to describe cold atom gases beyond mean-field.

At this point we note that there exist alternative ways of doing the
unitary transformation. Two examples, so called renormalization group
transformations, are {}``$V_{low-k}$'' \cite{Vlowk} and SRG (similarity
renormalization group) \cite{SRG}. SRG is frequently used in nuclear
physics \cite{SRG_in_nuclear_physics}. Another example, related
to SRG, is UCOM (unitary correlation operator method) \cite{UCOM}.
In future studies, it would be interesting to try alternative methods
for the problem at hand, but here we have focused on the Lee-Suzuki
transformation. Another study which applies the Lee-Suzuki transformation
on a non-nuclear system is \cite{morten}, in which the two-electron
problem in a quantum dot is examined. Also, in \cite{stetcu,alhassid-bertsch-fang}
the few-body problem of a trapped Fermi gas is investigated with techniques
similar to what is used here.

For an ultra-cold gas of neutral atoms, the interaction between the
particles can often be modeled by a short-range potential. We choose
to use a Gaussian distribution function parameterized by a range $\sigma$,
and a strength coefficient $g$. We here let the system be (quasi-)
two-dimensional. The Hamiltonian is 

\[
\hat{H}=\sum_{i=1}^{N}\frac{\hat{\mathbf{p}}_{i}^{2}}{2m}+\frac{1}{2}m\omega^{2}\mathbf{r}_{i}^{2}+\frac{1}{2}\sum_{j\neq i}^{N}g\frac{1}{2\pi\sigma^{2}}\exp(-\frac{|\mathbf{r}_{i}-\mathbf{r}_{j}|^{2}}{2\sigma^{2}}),\]
where $N$ is the number of particles and $m$ is the particle mass.
The chosen interaction is normalized so that it becomes a $\delta$-function
in the limit when $\sigma$ goes to zero, which also allows us to
avoid the mathematical difficulties of point-interactions in connection
to the diagonalization of the full many-body Hamiltonian \cite{delta}.
In a cold atomic gas, the strength coefficient $g$ is related to
the scattering length between the particles. The scattering length
can in some systems be experimentally tunable by Feshbach resonances
\cite{pitaevskii-stringari2003}. The typical length scale of the
system is the oscillator length, $\sqrt{\hbar/(m\omega)}$. The range
of the interaction in ultra-cold atomic gases is typically very short
compared to the wavelength of the particles; here we pick $\sigma=0.1\sqrt{\hbar/(m\omega)}$
for our calculations (in the following the oscillator length unit
is used). This parameter will be discussed in more detail later.

In two dimensions, the (s-wave) scattering length, $a$, is usually
defined to be a positive quantitity, see for example Ref. \cite{scatt_length_2D}.
A two-body system will have at least one bound state when the scattering
length is larger than the distance outside which the interaction potential
is zero. However, a purely repulsive short-range potential produces
a scattering length that is smaller than the range of the potential.
For such systems, an increased attraction (or reduced repulsion) would
decrease $a$ towards zero and an even stronger attraction would produce
a bound state and give an $a$ decreasing from infinity. The scattering
lengths corresponding to the repulsive interaction potential used
here can be calculated numerically, see table \ref{scatt_length_table}.

\begin{table}
\begin{tabular}{ccc}
Interaction strength & \,\,\,\,\,\,\,\,\,\,\,Range\,\,\,\,\,\,\,\,\,\,\, & Scattering length\tabularnewline
$g$ & $\sigma$ & $a$\tabularnewline
\hline
1 & 0.1 & 0.000283\tabularnewline
10 & 0.1 & 0.0871\tabularnewline
10 & 1 & 0.871\tabularnewline
\end{tabular}

\caption{\label{scatt_length_table}Numerically calculated s-wave scattering
lengths for some parameter choices of the interaction potential (all
quantities in oscillator units).}

\end{table}

The short-range interaction between the particles induces short-range
correlations, thus one would need a very large set of basis states
in order to accurately describe the wavefunction and to obtain a good
estimate of the energy of the system. However, one may perform a unitary
transformation of the Hamiltonian so that states within a given model
space become decoupled from the ones in the complementary excluded
space \cite{Lee-Suzuki,unitary-operator}. The finite model space
and the complementary infinite space can be defined by the projection
operators $\hat{P}$ and $\hat{Q}$, respectively. The original Hamiltonian
of the system is transformed to an effective Hamiltonian acting only
in a subspace of the complete Hilbert space, while preserving a subset
of the original eigenvalues. Thus, given the effective Hamiltonian,
the computational effort in finding the eigenvalues can be reduced.
However, even though the original Hamiltonian contains only one- and
two-body terms, the effective Hamiltonian will in general be an $N$-body
operator. So far, no approximation has been introduced, only an operator
transformation given by: \begin{equation}
\hat{H}_{\mathrm{eff}}=\frac{\hat{P}+\hat{P}\hat{\xi}^{\dagger}\hat{Q}}{\sqrt{\hat{P}+\hat{\xi}^{\dagger}\hat{\xi}}}\hat{H}_{\mathrm{{original}}}\frac{\hat{Q}\hat{\xi}\hat{P}+\hat{P}}{\sqrt{\hat{P}+\hat{\xi}^{\dagger}\hat{\xi}}},\label{eq:transformation}\end{equation}
 where $\hat{\xi}$ is an operator acting as a mapping between the
$P$- and $Q$-spaces, satisfying $\hat{\xi}=\hat{Q}\hat{\xi}\hat{P}$.
Note that the transformation~(\ref{eq:transformation}) is unitary
and yields an effective Hamiltonian that is energy-independent and
hermitian. $\hat{H}_{\mathrm{eff}}$ does not couple states in the
$P$-space with states in the $Q$-space.

Unfortunately, to find the mapping operator $\hat{\xi}$ one needs
the exact solution of the $N$-body problem, which is in itself the
final goal. Our simplest, yet nontrivial, approximation is to develop
a two-body effective Hamiltonian. The approximation consists in finding
$\hat{\xi}_{2}$ for the two-body problem and to compute an effective
Hamiltonian for the two-body system. By subtracting the one-body terms
(kinetic + potential energy) one thus obtains an effective two-body
interaction. This effective interaction is then used to construct
the $N$-body effective Hamiltonian that will now contain only one-
and two-body terms. Due to the approximation of the effective Hamiltonian,
the obtained energies will not be bound by the variational theorem.
This particular property of the current method is in contrast to the
situation encountered when using standard configuration interaction
calculations, or quantum Monte-Carlo approaches.

Although this approach can lead to good estimates of the energy eigenvalues,
the obtained eigenvectors are incorrect since they do not contain
components from the excluded $Q$-space. However, the eigenvectors
are often not interesting by themselves; only observables, expressed
as expectation values of physical operators, are relevant. Transformations
similar to the one performed on the Hamiltonian, can be applied to
other operators so that their expectation values can be obtained from
the $P$-space eigenvectors. In the present study, however, we restrict
the discussion to the many-body energy spectrum.

For bosons, the many-body basis states are permanents, with $N$ particles
distributed over the single-particle orbitals of the system. The truncation
of the infinite basis is performed in the following way: Neglecting
interactions, the state of lowest energy is the one with all $N$
bosons in the lowest orbital as defined by the confinement potential,
with energy $E=\hbar\omega N$. We incorporate in the $P$-space all
many-body states with an energy \[
E\leq\hbar\omega(N+\mathcal{N}_{\mathrm{max}}),\]
 so that $\mathcal{N}_{\mathrm{max}}$ is a parameter determining
the maximum allowed energy of particle-hole excitations from the state
of lowest energy. The number of included states increases rapidly
with $\mathcal{N}_{\mathrm{max}}$.

In addition, all possible combinations of two particles found in the
$N$-body model space ($P$) defines the restricted space of the two-body
system ($P_{2}$) that is used to compute the effective interaction.
Having defined $P_{2}$, the Hamiltonian for the complete two-body
space is transformed. In practice, the complete two-body space must
of course also be truncated. To this aim we take all two-body states
that can be constructed using the first 20 harmonic oscillator shells.

The original Hamiltonian preserves angular momentum ($L$), implying
that we can restrict the basis to states with a given value of $L$,
thus limiting its size. We confirm that the effective interaction
also preserves angular momentum (within numerical accuracy).

\begin{figure}
\begin{centering}
\includegraphics[%
  width=1\columnwidth,
  keepaspectratio]{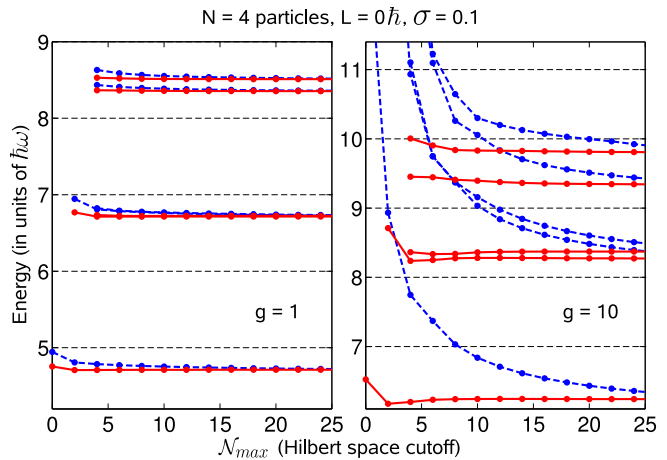}
\par\end{centering}

\caption{\label{fig:N4}Energies for a system of $N=4$ bosons and total angular
momentum $L=0\hbar$, for different cutoffs of the many-body Hilbert
space (parameterized by $\mathcal{N}_{\mathrm{max}}$). The range
of the interaction is $\sigma=0.1$, and two different strengths ($g$)
are shown. The \textcolor{blue}{\emph{blue dashed}} curves are the
results from standard configuration interaction calculations, while
the \textcolor{red}{\emph{red solid}} curves are energies obtained
using the effective interaction approach. While the standard calculations
require a large basis set to give a good energy estimate, the effective
interaction provides roughly the same answer with significantly less
computational effort. (Please note that for small $\mathcal{N}_{\mathrm{max}}$,
there are very few states in the $P$-space, and consequently only
a few eigenvalues can be obtained.)}

\end{figure}

First we use the method to calculate energies of a system with only
four particles, $N=4$, with an interaction range $\sigma=0.1$. Here,
we consider only states with zero angular momentum. For such a small
system, the standard configuration interaction method can be applied.
Figure \ref{fig:N4} shows results from both types of calculations.
For $g=1$, both methods give good energy estimates already at small
$\mathcal{N}_{\mathrm{max}}$. But for $g=10$ the standard calculations
require a large basis set to obtain a reasonably converged energy
estimate, while the effective interaction provides an answer with
significantly less computational effort. Typically, for an $\mathcal{N}_{\mathrm{max}}$
around 5, the basis size is of order $10$, while for $\mathcal{N}_{\mathrm{max}}$
around 25 it is of order $10^{4}$. These numbers depend strongly
on the system parameters $N$ and $L$, though.

Note that for $\mathcal{N}_{\mathrm{max}}=0$, the standard configuration
interaction calculation reduces to a perturbative calculation, where
the energy grows linearly with the strength of the interaction, $g$.
In addition, for sufficiently large $\mathcal{N}_{\mathrm{max}}$
the two methods will be equivalent, since the $P$-space is then actually
the full space.

The method works well both for the ground state energy and the excitations.
An interesting observation is that for $g=10$, at $\mathcal{N}_{\mathrm{max}}=6$
the standard calculation shows a degeneracy in the first excited state.
This effect is spurious, however, since the energies split up when
a larger basis is used. When using the effective interaction, this
false degeneracy does not occur.

\begin{figure}
\begin{centering}
\includegraphics[%
  width=1\columnwidth,
  keepaspectratio]{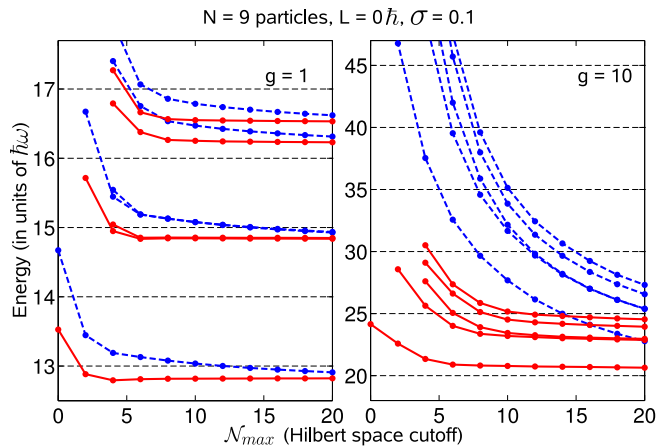}
\par\end{centering}

\caption{\label{fig:N9} Same as Fig.~\ref{fig:N4} but for $N=9$ bosons.
Note the rapid convergence of energies using the effective interaction
approach. For example, we note that in this case a calculation with
$\mathcal{N}_{\mathrm{max}}=6$ requires 12 basis states, while $\mathcal{N}_{\mathrm{max}}=20$
would correspond to 81097 states.}

\end{figure}

Let us now turn to systems with larger number of particles. Figure
\ref{fig:N9} shows calculations for $N=9$ bosons. In the case of
$g=1$ the effective interaction method again very rapidly produces
an accurate estimate of the energy. For $g=10$, where the correlations
in the systems are much stronger than for $g=1$, the energies obtained
when using effective interactions appear to have reached some plateau,
but still show a slow decrease with growing $\mathcal{N}_{\mathrm{max}}$.
In comparison, the energies obtained from the standard calculations
show a much slower convergence.

\begin{figure}
\begin{centering}
\includegraphics[%
  width=1\columnwidth,
  keepaspectratio]{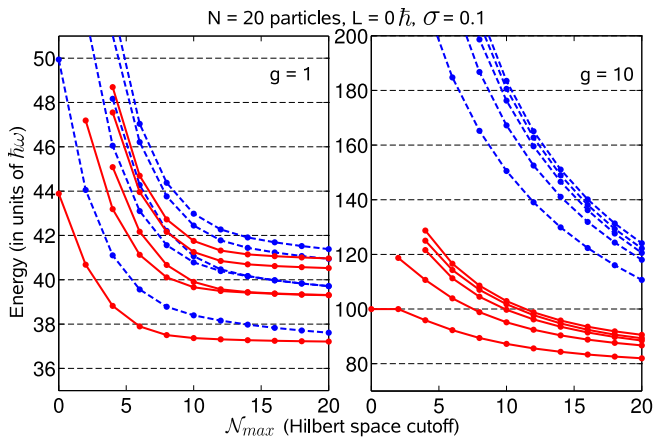}
\par\end{centering}

\caption{\label{fig:N20} Same as Fig~\ref{fig:N4} but for $N=20$ bosons.
For $g=10$, the energy obtained using effective interaction cannot
be said to be converged as a function of $\mathcal{N}_{\mathrm{max}}$.}

\end{figure}

The convergence behavior for an even larger system of $N=20$ particles
is shown in Fig.~\ref{fig:N20}. In the case of $g=1$ it is possible
to reach fairly converged results with both methods. When $g=10$
neither method is able to provide converged energies for the range
of $\mathcal{N}_{\mathrm{max}}$ considered here. An intuitive interpretation
would be that $\mathcal{N}_{\mathrm{max}}$ should be of the same
order as $N$, allowing about one extra unit of energy per particle,
so that every particle has some freedom to adjust in the system. Since
the number of basis states grows very rapidly with both $N$ and $\mathcal{N}_{\mathrm{max}}$,
it seems that the method is most suitable in the strongly correlated
regime; and makes it possible to study larger, but not much larger,
systems than with the standard method.

\begin{figure}
\begin{centering}
\includegraphics[%
  width=1\columnwidth,
  keepaspectratio]{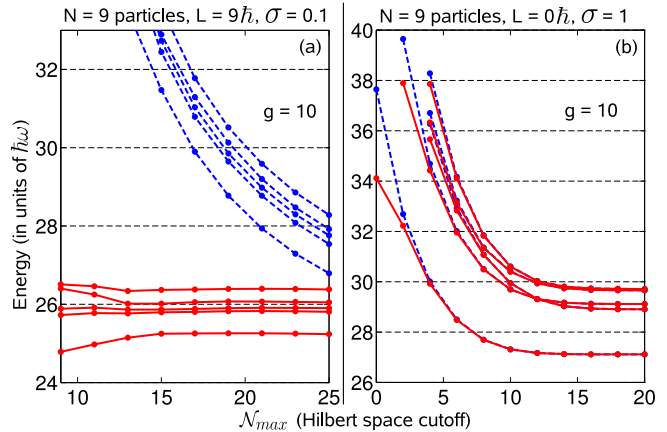}
\par\end{centering}

\caption{\label{fig:L9-and-sigma1}(a) Energy levels (in units of $\hbar\omega$)
for a system of $N=9$ bosons and angular momentum $L=9\hbar$ (in
this case there are no possible basis states for $\mathcal{N}_{\mathrm{max}}<9$).
(b) Energies for a system with $N=9$ and interaction range $\sigma=1$
($L=0\hbar$ states). See caption of figure \ref{fig:N4} for explanations.}

\end{figure}

All results shown so far involved the $L=0\hbar$ (non-rotating) states
of the different systems studied. However, the method is not restricted
to the non-rotating case. Figure~\ref{fig:L9-and-sigma1}(a) shows
the energies for a system of $N=9$ bosons, at angular momentum $L=9\hbar$.
Here, the energy is not a strictly decreasing function of $\mathcal{N}_{\mathrm{max}}$,
but aside from this the convergence is similar to that seen in figure
\ref{fig:N9}. An energy obtained with the standard configuration
interaction method is always an upper bound to the true value, since
it is a variational approach. As the size of the basis space is increased,
a better or equally good estimate is found. Calculations using an
approximated effective Hamiltonian, as in this study, are not variational
so higher-order terms may contribute with either sign to the energy.

The Lee-Suzuki approach was invented to handle the short-range correlations
in nuclei \cite{Lee-Suzuki,unitary-operator}, and the short range
of the interaction is known to be essential for the performance of
the method. For all results presented this far, we have set the range
parameter $\sigma=0.1$. Figure~\ref{fig:L9-and-sigma1}(b) shows
results with a larger range, $\sigma=1$, for a system with $N=9$
particles. Apart from some deviations for small $\mathcal{N}_{\mathrm{max}}$,
the effective interaction here does not give an improved convergence
rate compared to the standard calculations. This result suggests that
for $\sigma$ smaller than 0.1 the method would perform even better
than demonstrated in e.g. figure \ref{fig:N9}. However, the numerical
solution of the two-body problem becomes more difficult with decreasing
$\sigma$, and our present implementation prevents us from exploring
smaller $\sigma$.

As mentioned, the correct effective Hamiltonian would in general be
an $N$-body operator. It would be interesting to examine how inclusion
of e.g. an effective three-body interaction would affect the numerical
convergence, although the two-body approximation is exact in the limit
$N_{max}\rightarrow\infty$. As seen in our results, the obtained
energies typically converge rapidly as functions of $N_{max}$, implying
that an effective two-body Hamiltonian is sufficient in many situations.

To summarize, in order to calculate properties of cold, bosonic atomic
gases with strong short-range interactions between the particles,
we have employed a unitary transformation of the Hamiltonian. The
transformation is used to recover correlation effects which would
otherwise be lost within the heavily truncated basis space we consider.
The method is in practice a modification of the standard configuration
interaction approach. In many cases it produces a good estimate of
the energy, with a computational effort which is several orders of
magnitude smaller than that required by the standard method. The main
advantage, compared to many other approaches, is the accessibility
of the excitation spectrum with \textit{significantly} reduced numerical
effort.

\begin{acknowledgments}
We thank B.~Mottelson, C.~Pethick, D.~Pfannkuche, M.~Rontani and
H.~Smith for discussions regarding the use of pseudo-potentials in
exact diagonalization studies. S.Å. acknowledges B. Barrett for discussions.
This work was financed by the Swedish Research Council, the Swedish
Foundation for Strategic Research, and NordForsk. 
\end{acknowledgments}

\end{document}